\documentclass[a4paper,11pt]{article}
\usepackage{amsmath}
\usepackage{amssymb}
\usepackage{amsthm}
\usepackage{enumitem}
\usepackage{jinstpub} 
\usepackage{multirow}
\usepackage{graphicx}

\pdfoutput=1          

\title{\boldmath Pulse shape simulation of p-type Point Contact Germanium Detector for The MAJORONA DEMONSTRATOR Experiment }

\author[a,1]{Akash Pandey,\note{Corresponding author.}}
\author[a,]{D. Singh}

\author[a,b,2]{V. Singh,\note{Corresponding author.}}

\affiliation[a]{Banaras Hindu University, Department of Physics, Institute of Science\\Varanasi$-$221005, India}

\affiliation[b]{Central University of South Bihar, Physics Department, School of Physical and Chemical Sciences\\Gaya$-$824236, India}

\emailAdd{akash.pandey@bhu.ac.in}
\emailAdd{venkaz@yahoo.com}
\abstract{In the rare physics events search, liquid and solid detector plays an important role. Its analysis is based on the detection of excess events over the expected background or on the detection of an annual event rate  modulation. Germanium detectors sensitivities have been demonstrated as efficient means to probe Weakly Interacting Massive Particles (WIMPs). Germanium detectors having sensitivity less than 100 eV opens new platform for the investigation of neutrinos and dark matter physics. There are various working configuration available for germanium detectors such as semi-coaxial, true coaxial and point contact. The point contact germanium detectors have been widely used for dark matter and low energy neutrino searches because of their sub KeV energy threshold and low electronic noise~\cite{soma}. Pulse shape of P-type point contact (P-PC) germanium detector has been simulated in the present work and compared to the pulse shape of Majorana Demonstrator(MJD)~\cite{MJD}}

\begin{document}
\maketitle
\flushbottom

\section{Introduction}
\begin{figure}[!tbp]
  \centering
  \begin{minipage}[b]{0.57\textwidth}
   \fbox{ \includegraphics[width=\textwidth]{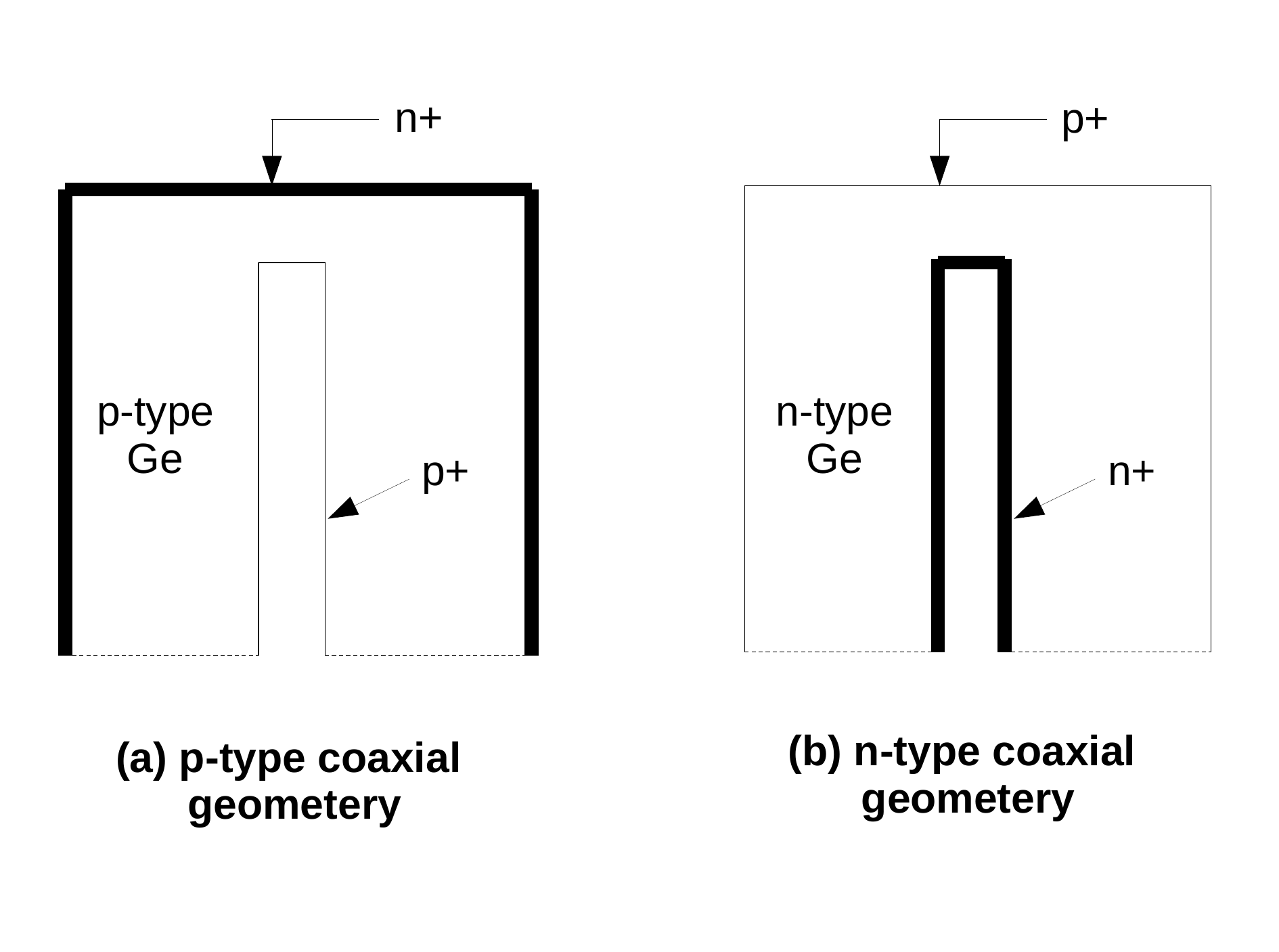}}
    \caption{Coaxial detector configuration}
    \label{coax}
  \end{minipage}
  \hfill
  \begin{minipage}[b]{0.57\textwidth}
   \fbox{ \includegraphics[width=\textwidth]{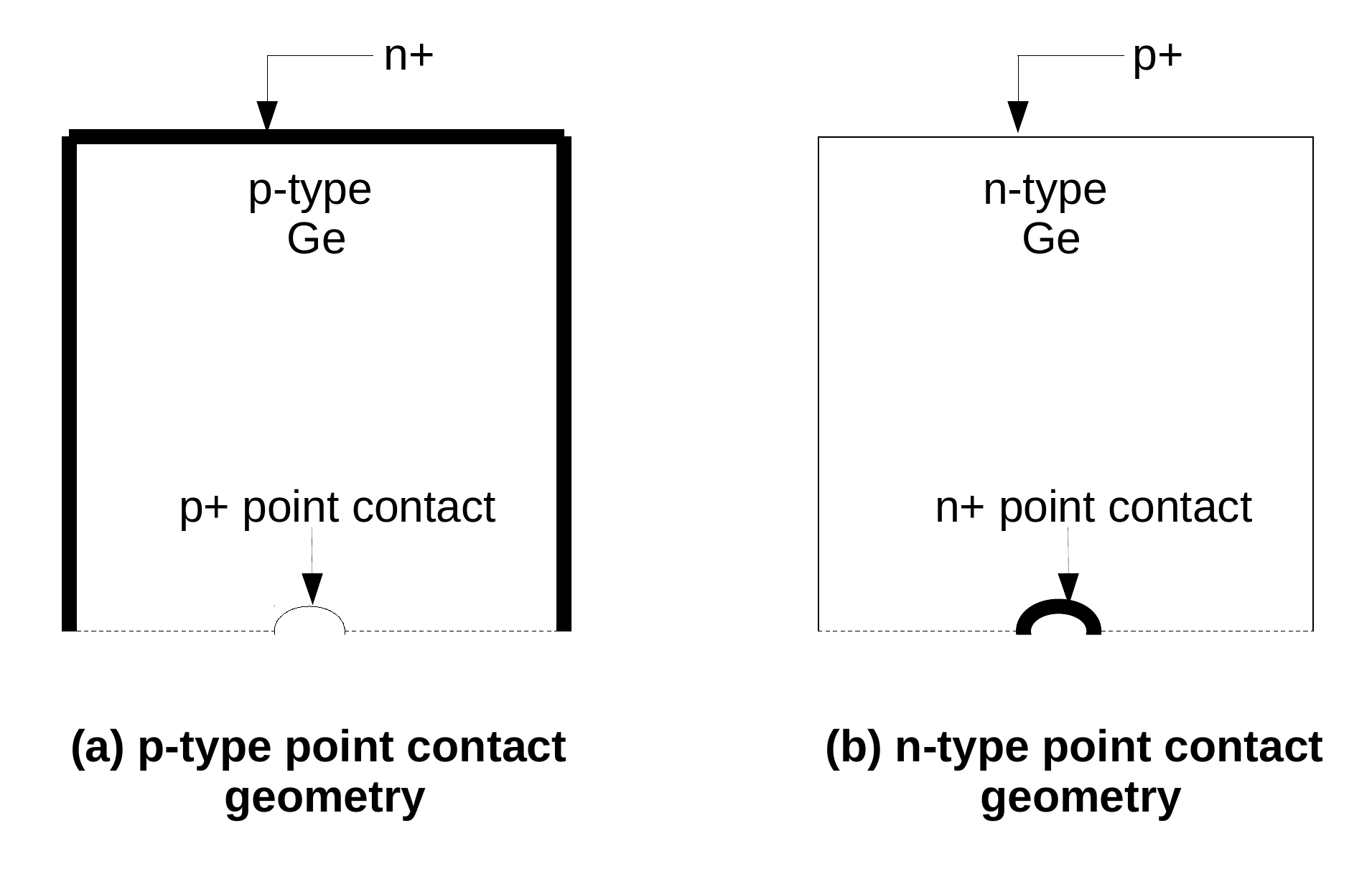}}
    \caption{Point contact detector configuration}
     \label{pc}
  \end{minipage}
\end{figure}

The most basic configuration of germanium detector is the coaxial configuration in which p-type and n-type both can be designed as shown in Fig.~\ref{coax}(a) and Fig.~\ref{coax}(b) . Outer dead layer (p+) of the n-type coaxial germanium detectors are made up of very thin layer ( 0.1 $\mu m$ - 1.0 $\mu m$ ) with boron implantation as shown in Fig.~\ref{coax}(b) . Thin layer allows low energy x-rays to the active region of the detector and give fragility to the detector. In the construction of p-type coaxial germanium detectors, unlike the n-type detector a thick outer dead layer (n+) ($\sim$ 0.5$ \mu$ $mm$) is fabricated with lithium ion diffusion as shown in Fig.~\ref{coax}(a). The thick designing of outer dead layer in the  p-type germanium detectors gives more robustness to the configuration as compared to the n-type germanium detectors.

Idea of utilizing point contact germanium detectors was firstly proposed in 1980s~\cite{luke} with n-type point contact germanium (nPCGe) with mass 800g in  which inner core of n-type coaxial  detector is replaced with a point contact as show in the Fig.~\ref{pc}(b). Designing of the p-type point contact germanium (pPCGe) detector can be achieve with the introducing the point contact structure over the coaxial shape of p-type coaxial germanium detector. Schematic diagram of p-type point contact germanium detector has been shown in the Fig.~\ref{pc}(a). 
Low threshold and low background are the key features of point contact detectors for better sensitivity in neutrino and dark matter physics~\cite{Barbeau}. They can be optimized amongst the various detectors such as n-type point contact , p-type point contact in order to achieve sub KeV energy threshold. These detectors are capable to achieve low output capacitance of the order of ($\sim$ pF). 
\par
Signal simulation is an important framework to understand the charge transport and charge induction takes place within the detector.
In the present article, simulation has been done to study the charge signal produced at the read out of the p-type point contact germanium (pPCGe) detector  with help of several numerical techniques.  The fundamental input for simulating the charge signal numerically is an electric field inside the detector.  The charge signal has been simulated for the pPCGe detector based on an arbitrary detector parameters ( e,g,. detector size, bias voltage, impurity concentration of p-type Ge crystal, dimension of point contact etc.) in order to demonstrate the physics behind the charge collection with the variation in time. Additionally, for the validation of the simulation package, charge signal has been also simulated for the pPCGe detector based on Majorona Demonstrator (MJD) experiment~\cite{MJD} . Majorona Demonstrator (MJD) is an experiment planed for the search of neutrinoless double-beta decay of (${}^{76}Ge)$ isotope with the designing from the mixed array of enriched germanium detectors in the p-type point contact germanium (pPCGe) configuration. 
\label{sec:intro}

\section{The p-type Point Contact Germanium (pPCGe) Detectors}
\subsection{Biasing configuration}
\begin{figure}
  \centering
\fbox{\includegraphics[width=0.4\linewidth]{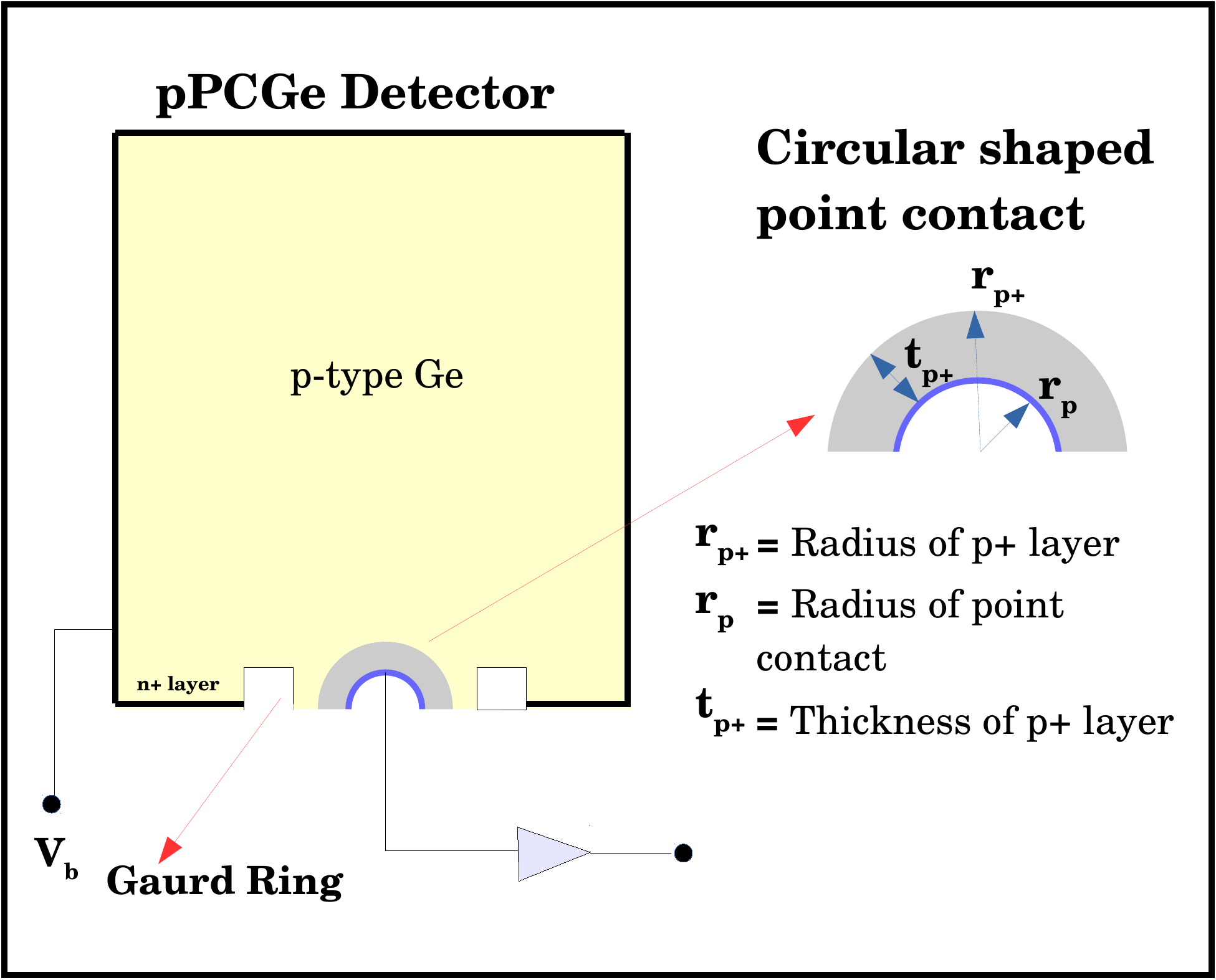}}
\caption{Schematic diagram ( sectional view) of P-PC detector having various components.}
\label{detector_geo}
\label{circular_point_contactt.pdf}

\end{figure}

The pPCGe detector is fully depleted with the high voltage followed by the reverse biasing configuration. Outer dead layer of the detector is a thick n+ layer fabricated using Li-ion diffusion and biased with high voltage. The circular shaped point contact, acting as a cathode, is grounded and act as readout as shown in  Fig.~\ref{detector_geo}. Grey region near the  vicinity of point contact (Fig.~\ref{detector_geo}) is a layer ( $\sim$ 300 nm ) composed of boron ion diffusion having concentration of the order of $10^{16} cm^{-3}$.

\subsection{Geometry of the point contact}
\begin{figure}[!tbp]
  \centering
  \begin{minipage}[b]{0.9\textwidth}
    \includegraphics[width=\textwidth]{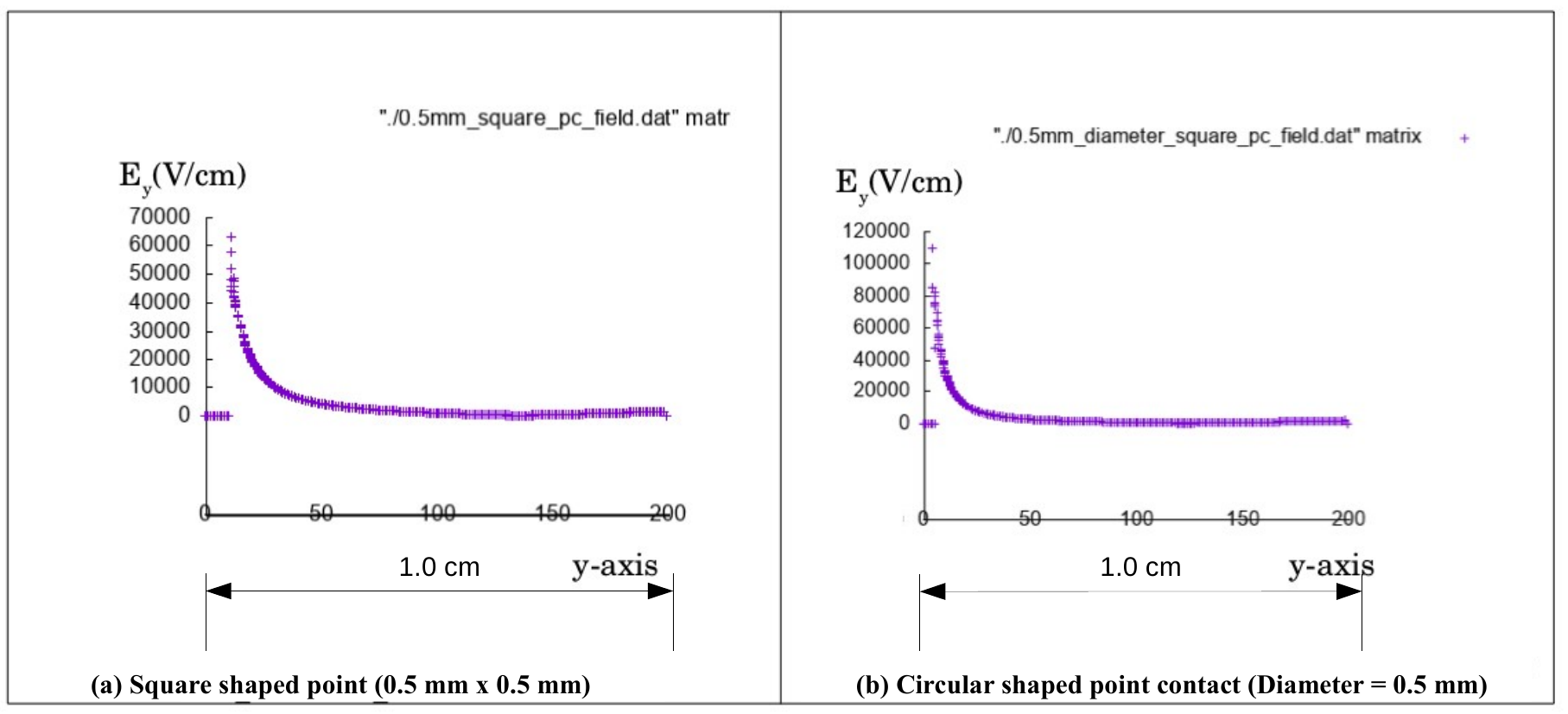}
    \caption{Effect of the point contact geometry on the electric field strength }
    \label{shape_effect}
  \end{minipage}
  \hfill
  \begin{minipage}[b]{0.9\textwidth}
    \includegraphics[width=\textwidth]{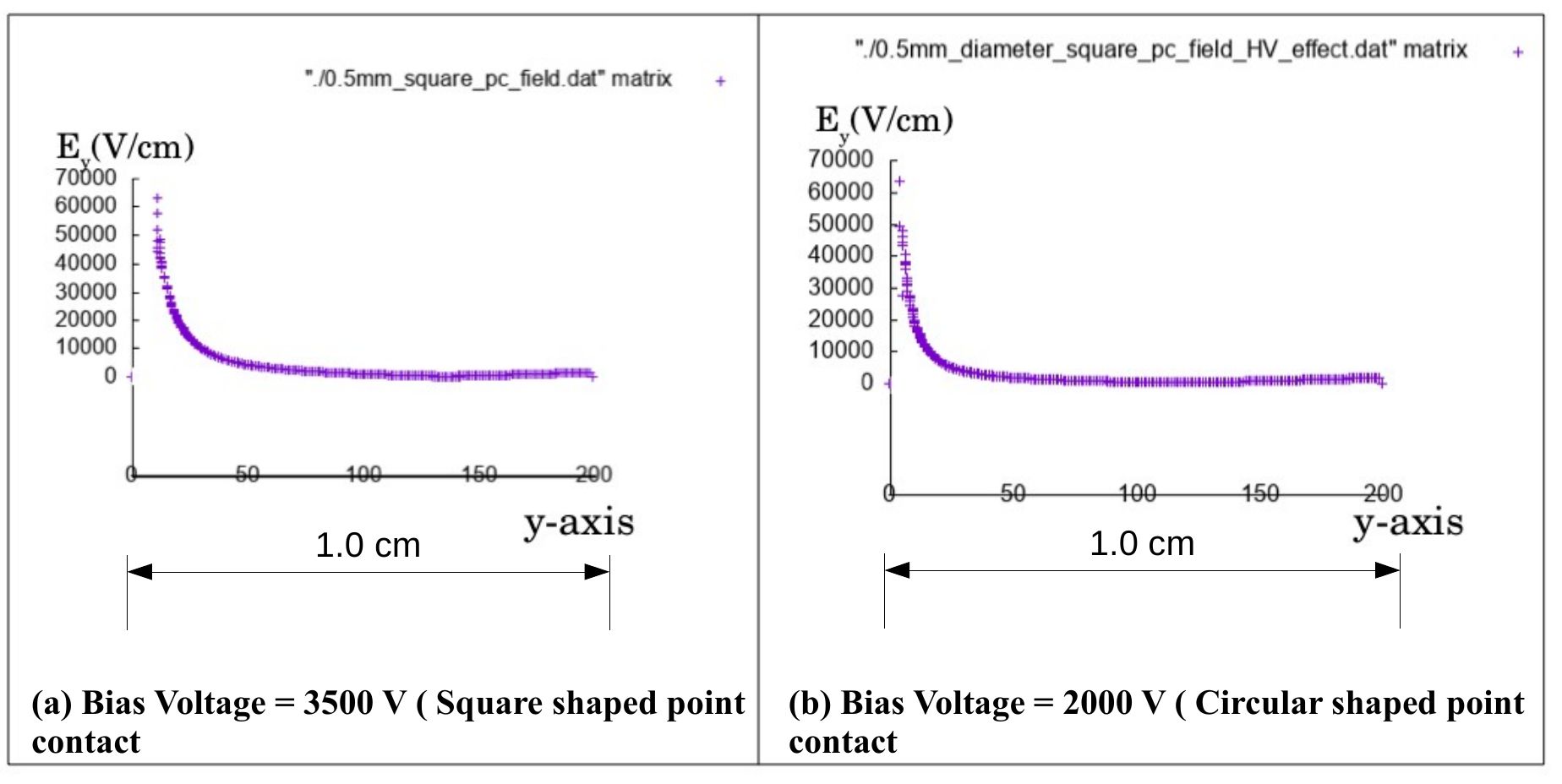}
    \caption{Effect of the point contact geometry on the biasing voltage}
    \label{adv_HV}
  \end{minipage}
\end{figure}

Circular point contact usually used in the most of the point contact germanium detectors. Advantage of the circular point contact design over any other shape is to obtained the higher electric field within the detector. Electric field has been simulated based on the square and circular shaped point contact in the Fig.~\ref{shape_effect}(a) and Fig.~\ref{shape_effect}(b) respectively and it has been observed that the strength of the maximum electric field for circular shaped point contact detector ($\sim 10^{5}$ V/cm ) is more than that of square shaped point contact ($\sim 10^{4}$ V/cm ) detector. \par Another important key feature in using the circular point contact shape is to minimize the value of detector bias voltage.
Same magnitude of maximum electric field can be obtained at relatively lower biasing voltage for the detector having circular point contact shape than that of squared point contact shape as shown in the Fig.~\ref{adv_HV}(a) and Fig.~\ref{adv_HV}(b) respectively. Biasing voltage plays an important role in the leakage current of the germanium detectors. It has been experimentally measured~\cite{wei} that the leakage current within the germanium detectors will go on increasing with applying higher magnitude of bias voltage.  As a consequence, the leakage current due to higher value of bias voltage can be minimized when the detector have been configured under the circular point contact geometry.

\subsection{ Designing of the Guard rings }
\begin{figure}[!tbp]
  \centering
  \begin{minipage}[b]{0.9\textwidth}
    \includegraphics[width=\textwidth]{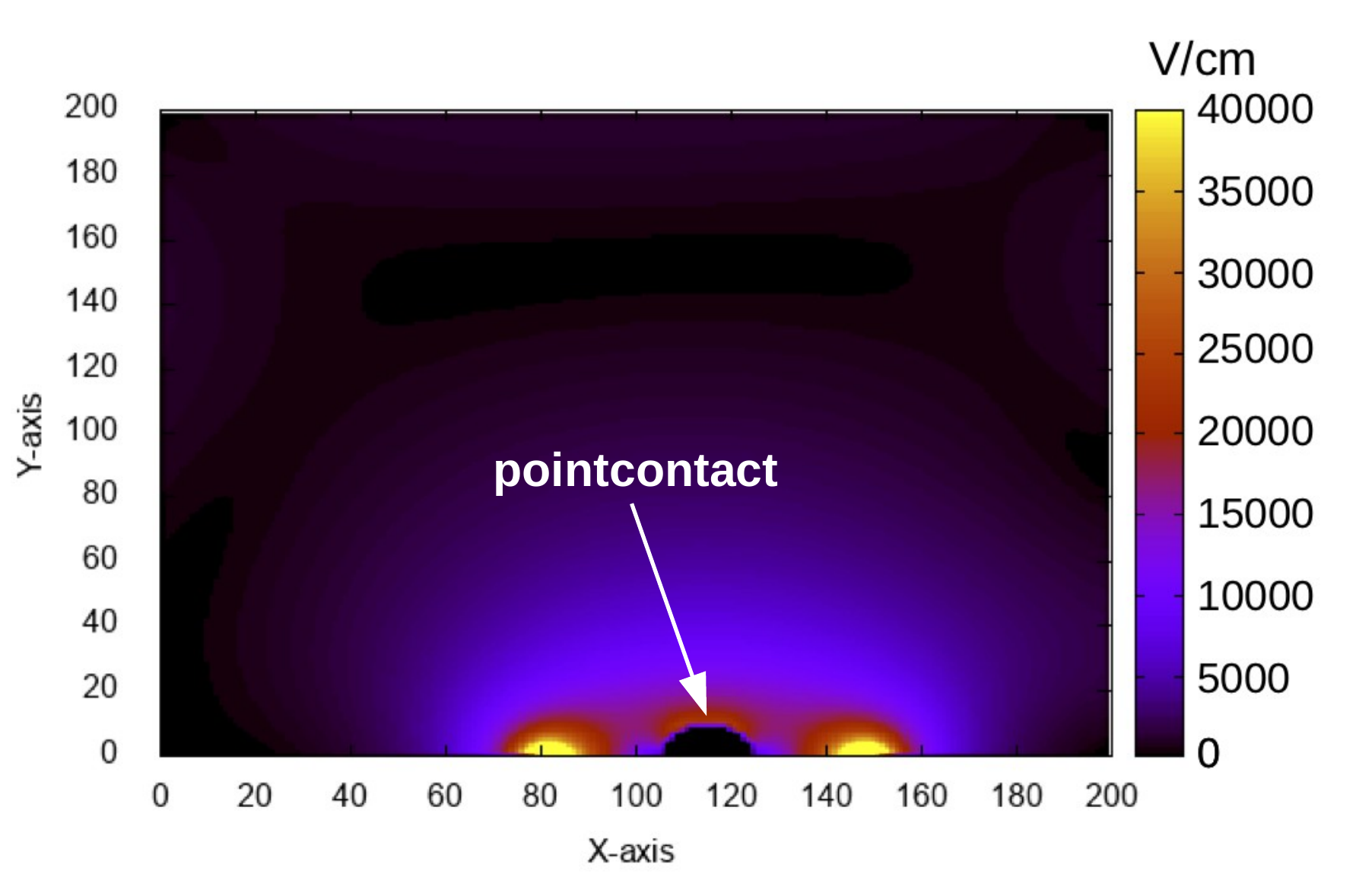}
    \caption{Variation of the electric field in the absence of guard ring }
    \label{no_gr}
  \end{minipage}
  \hfill
  \begin{minipage}[b]{0.9\textwidth}
    \includegraphics[width=\textwidth]{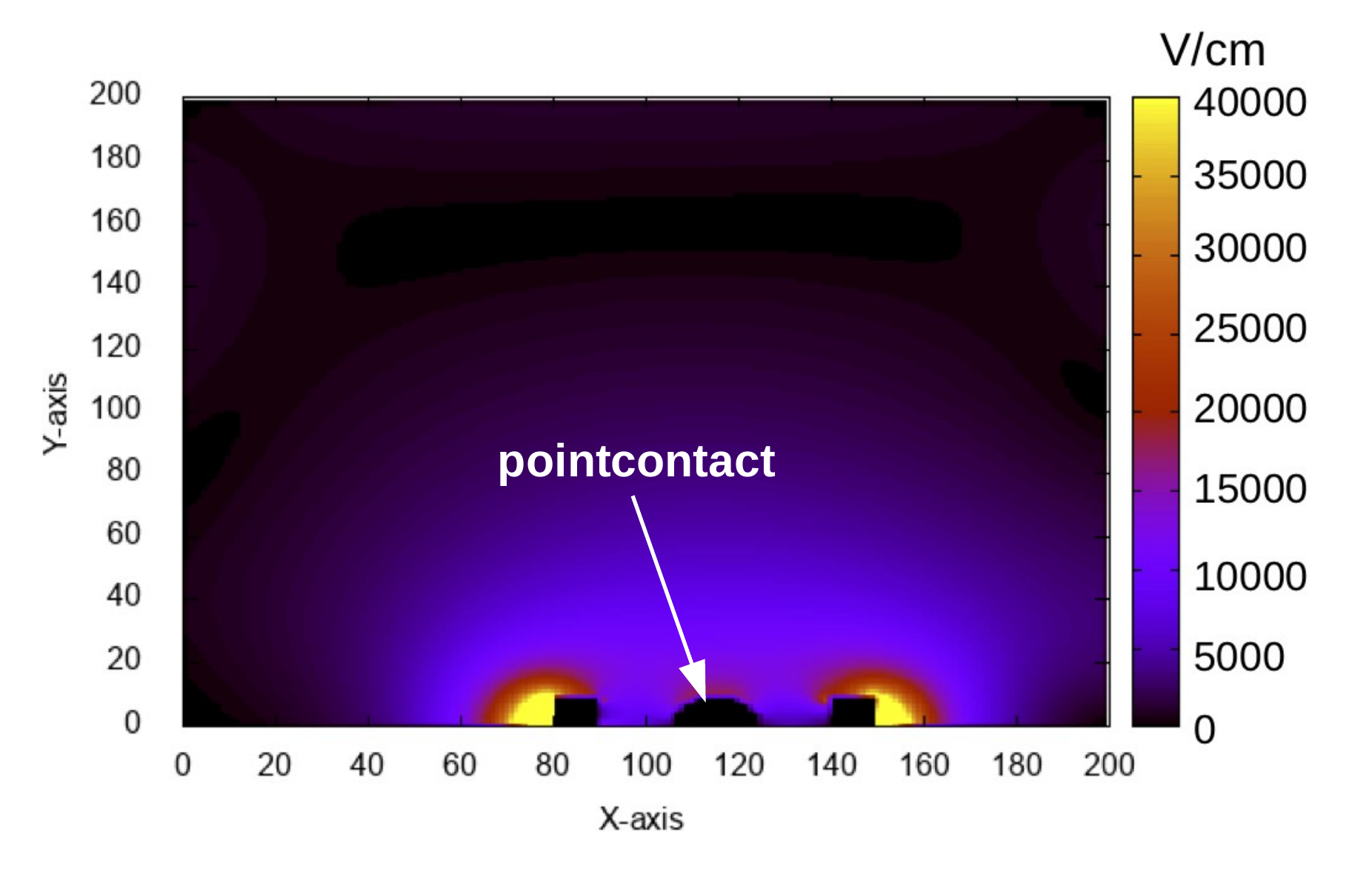}
    \caption{Variation of the electric field in the presence of guard ring}
    \label{gr}
  \end{minipage}
\end{figure}

Additional components of the detector are guard ring which protects the point contact regime from the spread of electric field to overcome the breakdown or short circuit of the point contact . Fig.~\ref{no_gr} shows the variation of electric field lines in the absence of guard rings in which field lines are spread over the point contact due to n+ biased layer. Upon introduction of guard rings(Fig.~\ref{gr} near the vicinity of point contact, spread of field lines is limited and point contact is protected from the breakdown/short circuit  due to n+ biased layer. Amorphous germanium has been spattered, having concentration usually of the order of $\sim 10^{14} cm^{-3} $, in the vicinity of guard ring in the same way as the boron layer diffusion done in the vicinity of the point contact.
\label{sec:heat_transfer}

\section{Simulation techinques}

\subsection{ Layout of Charge signal simulation}
\begin{figure}
  \centering
\fbox{\includegraphics[width=0.45\linewidth]{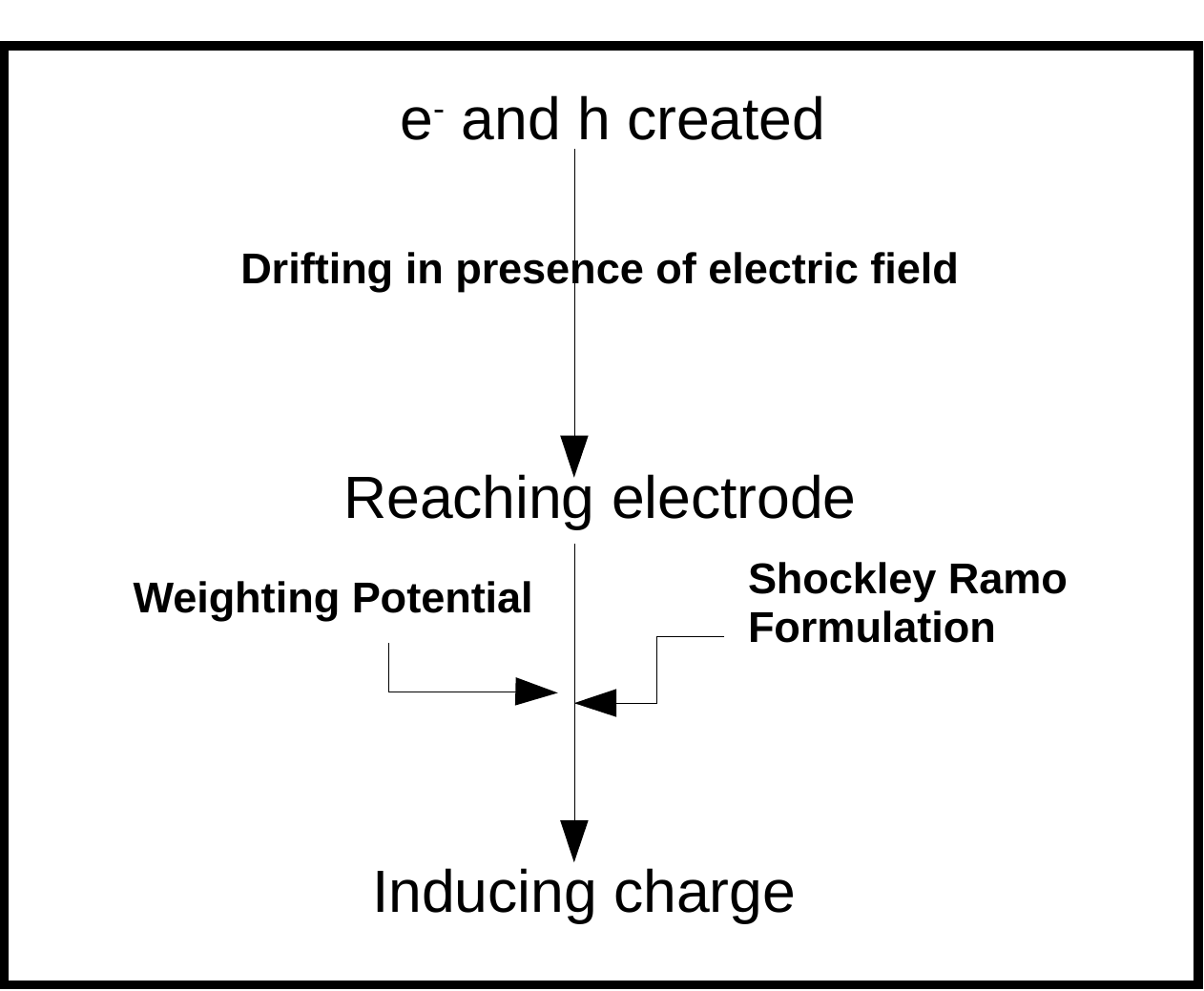}}
\caption{Flow chart of charge signal simulation}
\label{algo}
\end{figure}

The road map has been sketched in the Fig.~\ref{algo} to obtained the simulated charge signal pulse for the pPCGe detector. 
 Application of Numerical techniques used in each step of the layout has been discussed in the upcoming sections. 
\subsection{Electric Field simulation}
\subsubsection{Numerical Technique (Finite Difference Method)}
In order to simulate the 2D electric potential profile inside the detector , 2D Poisson's equation has been used. 2D Poisson's equation contains partial derivatives of electric potential ( $\phi(x,y)$ ) with respect to variable x and y respectively as given below:
     \begin{equation}
    \frac{\partial^{2} \phi(x,y)}{\partial x^2} +  \frac{\partial^{2} \phi(x,y)}{\partial y^2} =- \frac{ \rho}{\epsilon} 
      \label{RVsNPart}
    \end{equation}
    
       \begin{equation}
    or, \nabla^{2}\phi(x,y) = - \frac{ \rho}{\epsilon} 
      \label{RVsNPart}
    \end{equation}
    where $\rho$ is the impurity charge density, which is assumed to be uniform.
    \par
    Electric field has been calculated from the negative gradient of the electric potential as given below:
     
       \begin{equation}
  E = -\nabla \phi(x,y)
      \label{RVsNPart}
    \end{equation}

\begin{figure}
\includegraphics[width=0.8\linewidth]{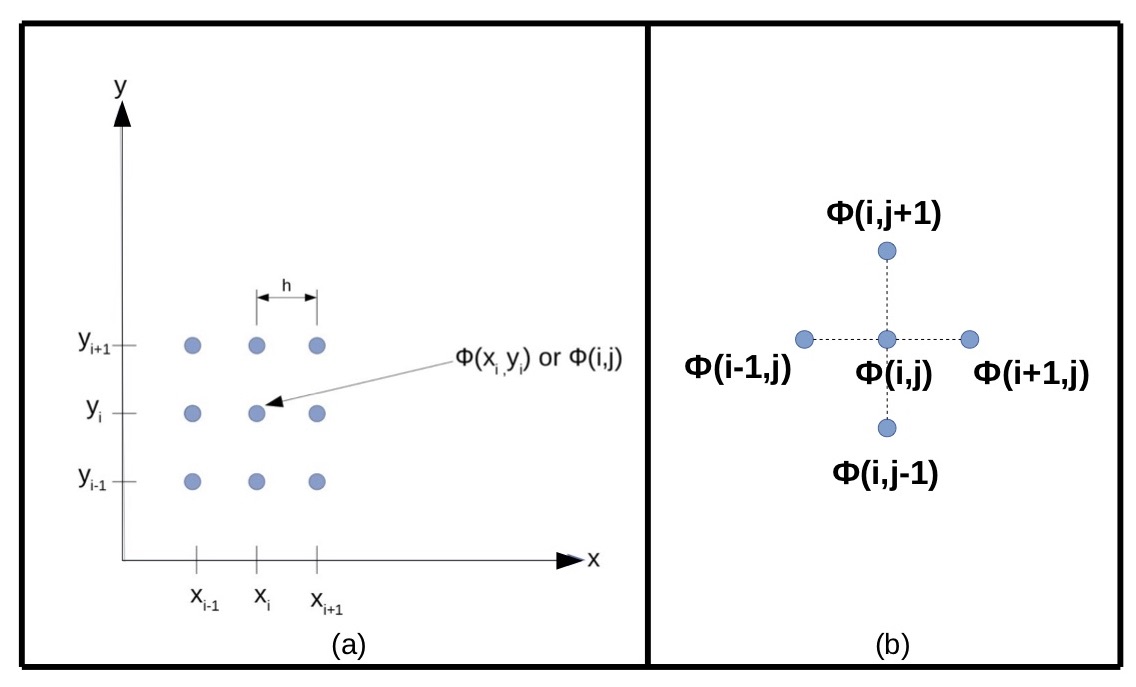}
\caption{(a)Grid Meshing in 2D (b) Five-point stencil}
\label{gridm}
\end{figure}

    Analytical solution of 2D poisson's is very tidy and needs lot of approximations. Therefore, the best suitable approach to obtain the solution is Numerical Techniques.
    There are several numerical techniques available to solve the 2D partial differential equations such as Finite Element Method, Finite Difference Method etc, .
    In the present work, Finite Difference Method (FDM) has been used to solve the 2D Poisson's equation for electric potential inside the detector.  
     In this method (x,y) domain of the detector has been meshed as shown in the Fig.~\ref{gridm}(a), into uniform grid of spatial points separated by $h$ at which
    electric potential will be  sampled. Electric potential on each and every single point in the detector domain is contributed by the four points in its neighborhood as shown in Fig.~\ref{gridm}(b). 
Formation of grids are obtained by :
  \begin{equation}
    x_{i} = ih
   \label{RVsNPart}
  \end{equation}
  \begin{equation}
   y_{j} = jh
     \label{RVsNPart}
 \end{equation}
 where i and j are integers.
  Therefore, the partial derivatives of poisson equation can be replaced by finite differences as given below:
     \begin{equation}
    \frac{\partial^{2} \phi(x,y)}{\partial x^2} = \frac{\phi(i-1,j)-2\phi(i,j)+\phi(i+1,j)}{h^{2}}
     \label{RVsNPart}
    \end{equation}
Similarly, 
\begin{equation}
    \frac{\partial^{2} \phi(x,y)}{\partial y^2} = \frac{\phi(i,j+1)-2\phi(i,j)+\phi(i,j-1)}{h^{2}}
     \label{RVsNPart}
    \end{equation}

\begin{figure}
 \centering
\includegraphics[width=0.5\linewidth]{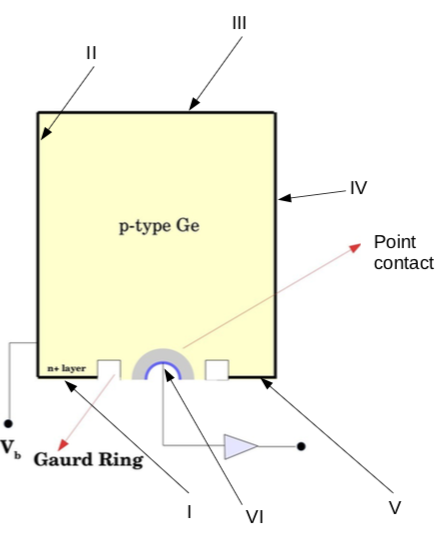}

\caption{Drichilet boundaries for different surfaces of the pPCGe detector.}
\label{bound}
\end{figure}

Following inputs has been used to solve the Poisson equation in order to determine the potential inside the detector region:

\textbf  {1.} Boundary conditions depends on the detector geometry. Such 
 conditions defines the shape and size of detector, point contact and gaurd rings. There are several types of boundary conditions exists for solving partial differential equations.
 In the present scenario, Drichilet boundary condition has been used. These boundary conditions are applied to the electric potentials $\phi(x,y)_{I}$ , $\phi(x,y)_{II}$ , $\phi(x,y)_{III}$ , $\phi(x,y)_{IV}$ , $\phi(x,y)_{V}$  on the n+ surface of the detector namely $I , II , III , IV , V$  respectively. All of these surfaces excluding $VI$ ( point contact) has been biased to the high voltage as shown in the Fig.~\ref{bound}. The surface $VI$ is the geometry of the point contact of the detector which serves as the read out and connected to the ground (zero potential) in order to work the detector in the reverse bias mode .

 \textbf  {2.} Ge parameters, i,e,.Impurity concentration (number density of acceptor atoms in case of p-type Ge crystal), Dielectric constant .
           
\textbf  {3.} Reverse bias voltage ( high voltage) .

 \subsubsection{Iteration method (Gauss-Seidel Method)}
	 Iterative method (Gauss-Seidel Iteration) has been used to simulate potential $\phi(i,j)$. 
  Gauss-Seidel method, is a modified version of Jacobi method also called as Successive Displacement Method, has been developed by Carl Friedrich Gauss and Philipp Ludwig von Seidel.
  Unlike the Jacobi method in which the values in each iteration obtained from the previous step, Gauss-Seidel method utilizes the value from the latest obtained value. 
This method is called as successive displacement method because the second unknown has been calculated from the first unknown and third unknown has been obtained from the first as well as second unknown in the current iteration.
Similar procedure has been repeated for the further iteration cycle. Continuity of the iteration cycle breaks down when pre declared accuracy achieved. In the present simulation the value of accuracy has been set to the order of ($\sim10^{-8}$).

\subsection{Pulse Shape simulation}

\subsubsection{Numerical Technique (Shockley-Ramo Method)}

\begin{figure}
          \centering
\fbox{\includegraphics[width=0.32\linewidth]{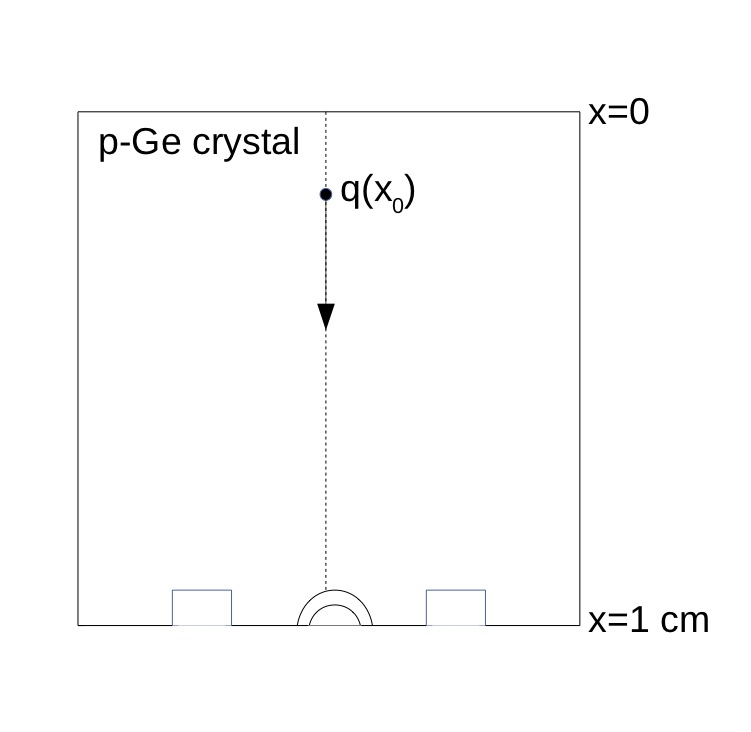}}
\caption{Motion of charge towards the read out (point contact)}
\label{geo.png}

\end{figure}

Shockley-Ramo theoram\cite{gatti} \cite{he} has been used to simulate the induce charge and induce current at the read out (point contact) of the pPCGe detetctor.
Based on \textbf {Shockley-Ramo theoram}, the induced charge $Q$ on the electrode by a point charge $q$ ( Fig.~\ref{geo.png}) located at position $x_{0}$ is given by:
     \begin{equation}
   Q_{q} = - q \psi(\overrightarrow x_{0})
    \label{RVsNPart}
      \end{equation}
      where $\psi(\overrightarrow x)$ is weighting potential defined as;
    \begin{equation}
   \nabla^{2}\psi(\overrightarrow x) = 0
    \label{RVsNPart}
      \end{equation}
      subjected to condition that potential is unity on read out electrode and else where it is zero.
      And field corresponding to weighting potential is called as weighting field given as;
        \begin{equation}
 \overrightarrow  E_{\psi(\overrightarrow x)} = -\nabla^{}\psi(\overrightarrow x) 
    \label{RVsNPart}
      \end{equation}
      
      Induce current $I_{q}.  $ to the electrode is given by:
\begin{equation}
I_{q} = \frac {dQ_{q}}{dt} = -q \big(\frac{\partial \psi}{\partial x_{0}}\frac{\partial x_{0}}{\partial t} +\frac{\partial \psi}{\partial y_{0}}\frac{\partial y_{0}}{\partial t}+\frac{\partial \psi}{\partial z_{0}}\frac{\partial z_{0}}{\partial t}\big)
    \label{RVsNPart}
      \end{equation}
     \begin{equation}
     or, I = q \overrightarrow E_{\psi}(\overrightarrow x_{0}) \overrightarrow v_{drift}
    \label{RVsNPart}
      \end{equation}
      
\subsubsection{Iteration used}

When the radiation entered in the bulk of the the detector, energy has been deposited in the detector volume. As a consequence of energy deposition , charge carriers (electrons and holes) has been created in the detector volume and starts drifting in the presence of electric field . Drifting of charge carriers has been guided from the applied electric field of the pPCGe detector.
\par
If the charge particle created at the initial position $x_{0}$ , the next position of the charge particle has been calculated as follows:
\begin{equation}
x = x_{0} + V_{d}\delta t
    \label{RVsNPart}
      \end{equation}
    where, $V_{d}$ is the drift velocity of charge carrier and $\delta t$ is the time step taken as 1 nano seconds in the present simulation.
    Each iteration cycle calculates the next position from the previous position of the charge carrier as the input keeping the same time step. 
    Iteration cycle remains continue till the charge carrier reached point contact.

\label{sec:MLI-Tech}

\section{Results and discussions}

\subsection{Variation of Weighting potential and Electric Field in P-PC germanium detector}
\begin{figure}[!tbp]
  \centering
  \begin{minipage}[b]{0.45\textwidth}
 \includegraphics[width=\textwidth]{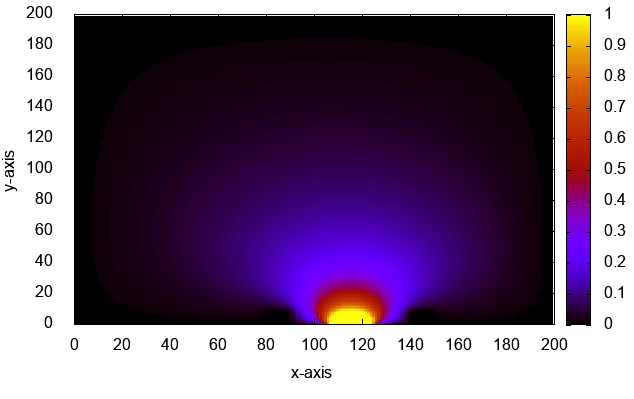}
    \caption{A 2D contour of weighting potential in pPCGe detector}
    \label{weighting_potential.png}
  \end{minipage}
  \hfill
  \begin{minipage}[b]{0.45\textwidth}
\includegraphics[width=\textwidth]{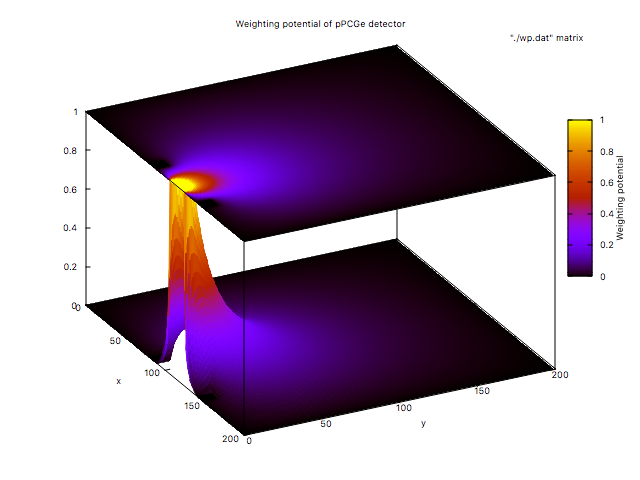}
    \caption{A 3D map of weighting potential  pPCGe detector}
    \label{weightingpotential.png}
  \end{minipage}
\end{figure}

\begin{figure}
\centering
\includegraphics[width=0.55\linewidth]{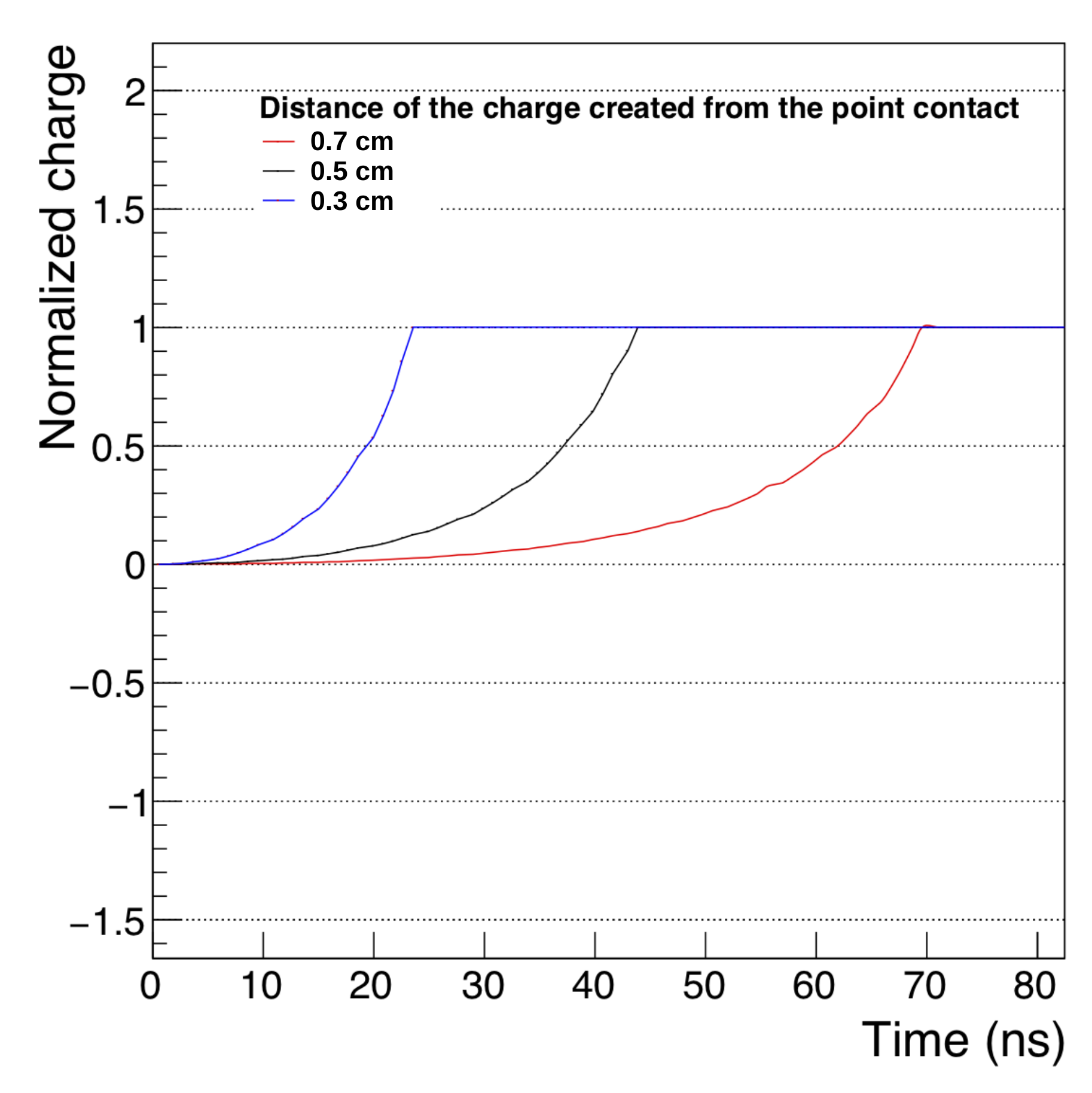}
\caption{Charge signal due to the charge carrier created at the different distances in the detector from the point contact}
\label{induce_charge_distance_effect}
\end{figure}

Consequence of Shockley-Ramo theorem simulate the variation of weighting potential within the pPCGe detector in the 2D and 3D maps as shown in Fig.~\ref{weighting_potential.png} and Fig.~\ref{weightingpotential.png}, respectively.
In the vicinity of point contact, strength of potential is found to be higher than the area which is away from it. Consequently, the charge carrier feels more drifting near the point contact and slows down as moving away from it. Therefore, the charge carriers, which are created in the nearby regime of the point contact, will reach the point contact relatively more quickly. 
 As a result, the rise time of charge signal found to be decreasing as the distance between point contact and the position of the charge carrier decreasing. Charge signals has been simulated for the different positions inside the detector and has been compared as shown in the Fig.~\ref{induce_charge_distance_effect}.
\par
\begin{figure}[!tbp]
  \centering
  \begin{minipage}[b]{0.45\textwidth}
    \includegraphics[width=\textwidth]{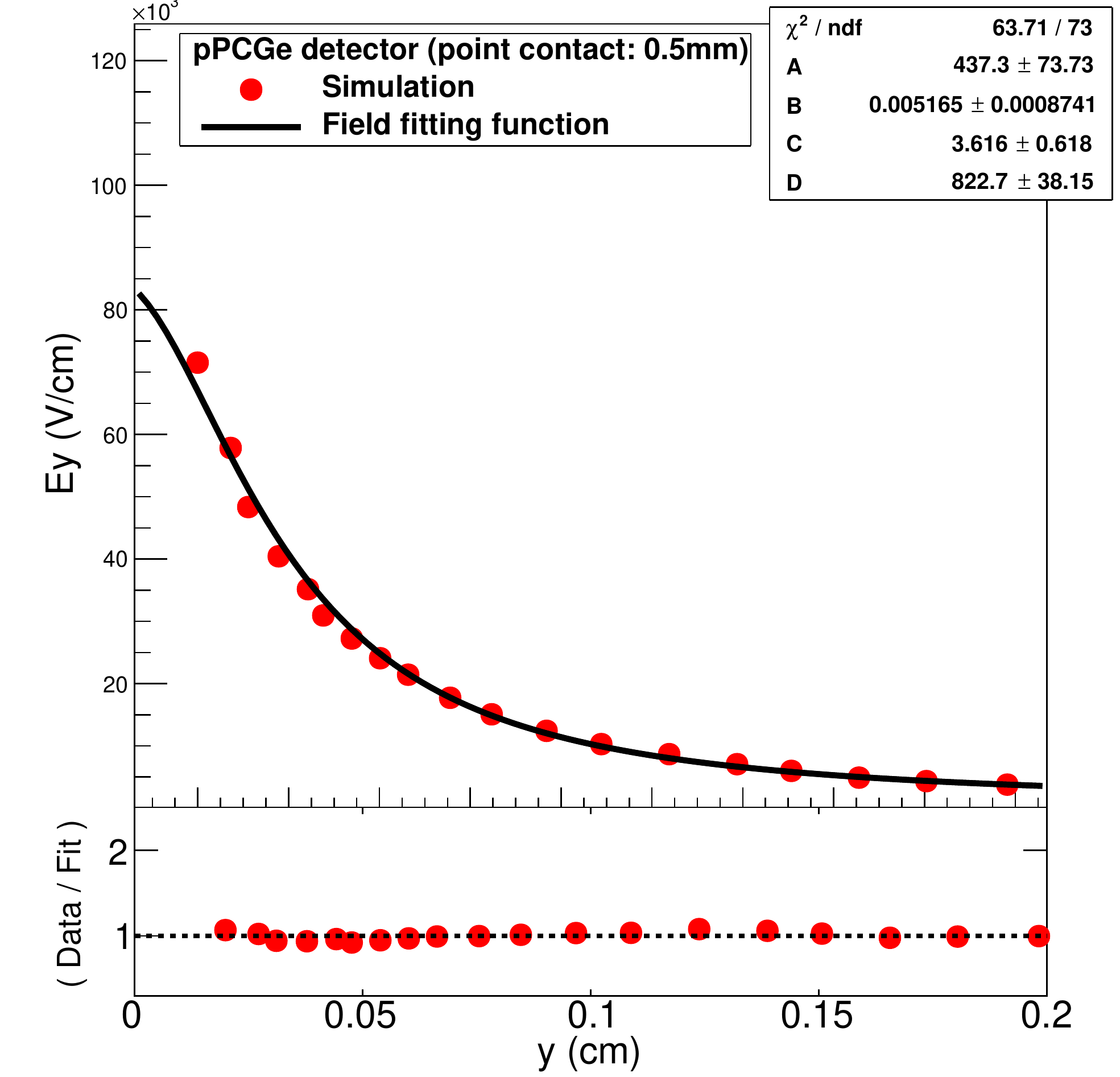}
    \caption{Variation of electric field in the pPCGe detector }
    \label{pcge_field}
  \end{minipage}
  \hfill
  \begin{minipage}[b]{0.5\textwidth}
\includegraphics[width=\textwidth]{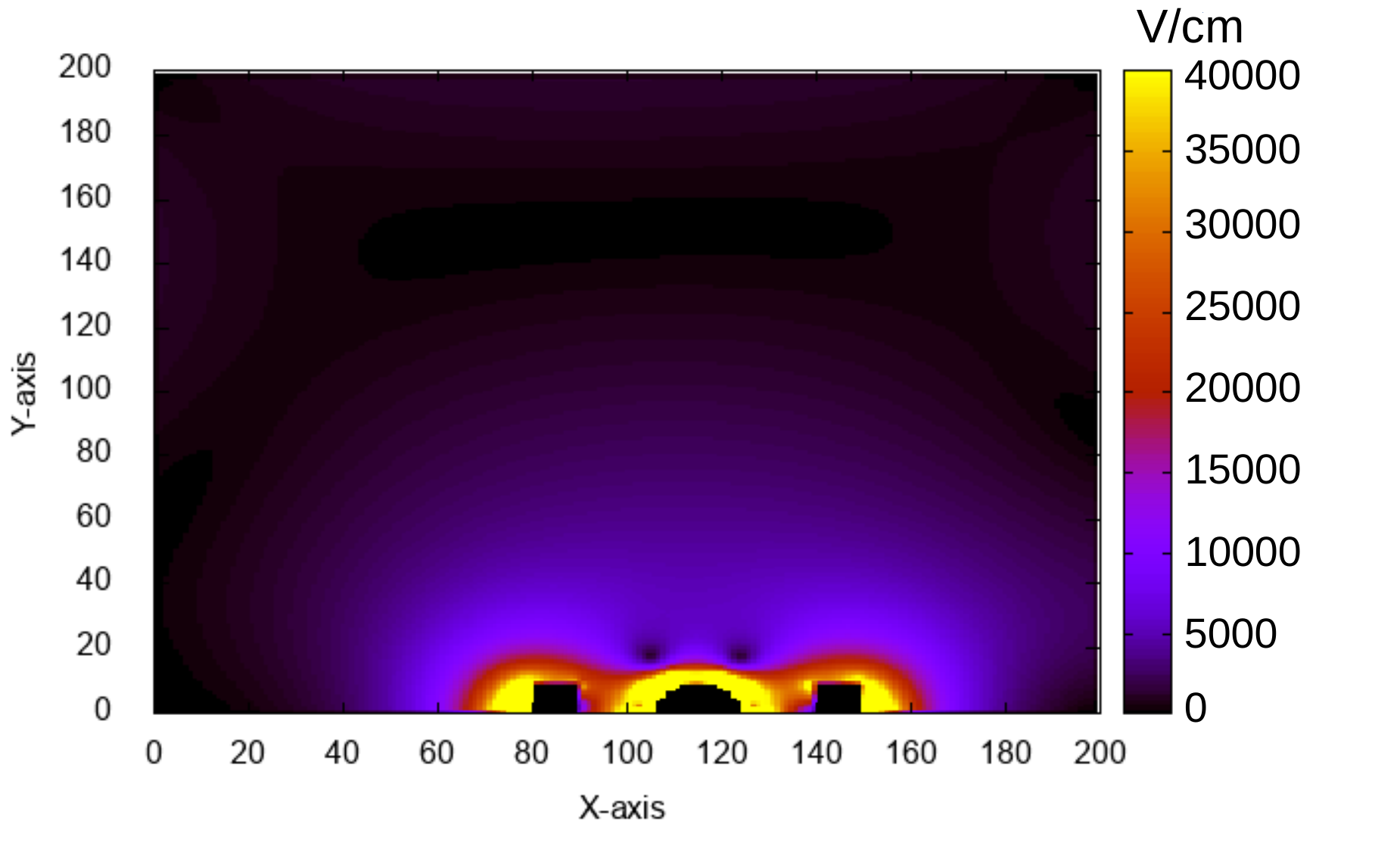}
    \caption{ A 2D contour plot of electric field in the pPCGe detector}
    \label{gaurd_ring_pc_boron_layer}
  \end{minipage}
\end{figure}

Numerical values of electric field, simulated from Finite Difference Method, has been fitted and its variation have been shown in  Fig.~\ref{pcge_field} as a function of distance from the point contact of the detector. Maximum electric field ($\sim 10^{4} V/cm$) has been observed in the close vicinity of point contact.
The contour map of electric field (Fig.~\ref{gaurd_ring_pc_boron_layer} ) shows that the region near the close vicinity of point contact have been majorly contributed with electric field and field strength becomes debilitated as moving away from the point contact.

\subsection{Drift velocity profile for germanium crystal at cryogenic temperature }

Drift velocity of charge carrier inside the detector is electric field dependent parameter. Linear variation of drift velocity with applied electric field is well known at the low field.
Where as in the high field region, change in drift velocity with electric field is small and saturated at the higher value of electric field beyond the specific point~\cite{Mei}~\cite{v_sat}.  Saturation of drift velocity can be described as:
 \begin{equation}
 V_{d} = \mu(E) E
 \label{RVsNPart}
      \end{equation}
      
      Where $\mu(E)$ is the field dependent mobility.
      This relation can be simplified into an empirical expression as :
      \begin{equation}
V_{d} = \frac{\mu_{0}E}{1 + \frac{E}{E_{sat}}}
    \label{RVsNPart}
      \end{equation}
     where $\mu_{0}$ is the mobility in the absence of field~\cite{zero_field} ; $E_{sat} = \frac{v_{sat}}{\mu_{0}}$, where $v_{sat}$ is known as the saturation drift velocity~\cite{v_sat}
     
     Therefore, the drift velocity ($v_{d}$) can be expressed as ;
       \begin{equation}
V_{d} = \frac{\mu_{0}E}{1+ \frac{\mu_{0}E}{v_{sat}}}
    \label{RVsNPart}
      \end{equation}
      Saturation velocity ($v_{sat}$) can be calculated using empirical formulae\cite{v_sat} given below:
      herefore, the drift velocity ($v_{d}$) can be expressed as ;
       \begin{equation}
V_{sat} = \frac{v^{sat}_{300}}{1 - A_{\nu} + A_{\nu}(\frac{T}{300})}
    \label{RVsNPart}    
      \end{equation}
      
      Velocity saturation model\cite{v_sat} gives the parameter values at temperature  300 K as: $v^{sat}_{300} = 0.7 x 10^{7} cm/s$ for electrons $0.63 x 10^{7} cm/s$, $A_{\nu } = 0.55 $ for electrons, $A_{\nu  } = 0.6 $1or holes

\par
\begin{figure}[!tbp]
  \centering
  \begin{minipage}[b]{0.5\textwidth}
    \includegraphics[width=\textwidth]{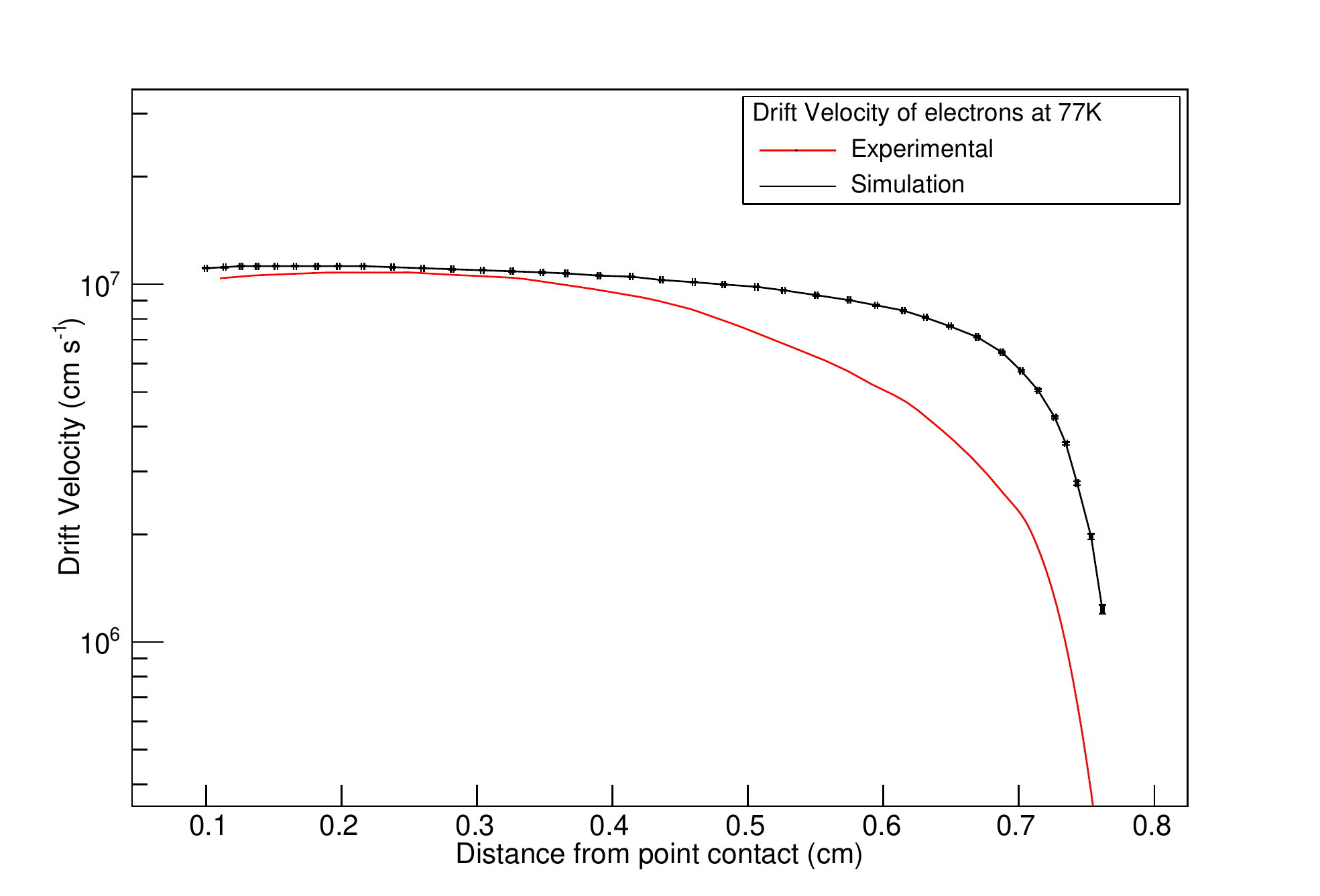}
    \caption{Drift veocity profile of electron within pCGe detector at 77 k}
    \label{drfit_electron_exp_sim}
  \end{minipage}
  \hfill
  \begin{minipage}[b]{0.5\textwidth}
    \includegraphics[width=\textwidth]{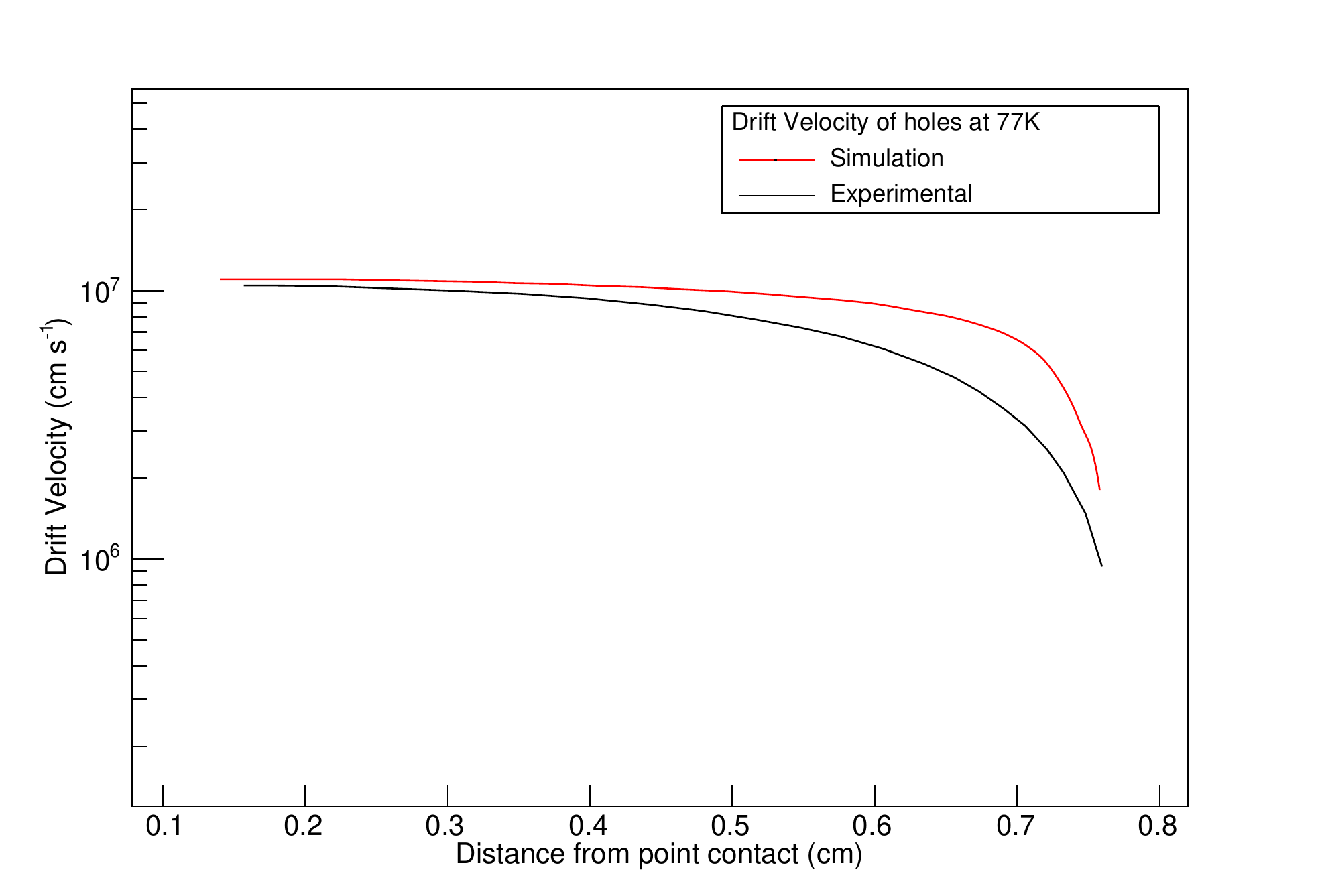}
    \caption{Drift veocity profile of hole within pCGe detector at 77 k}
     \label{drfit_hole_exp_sim}
  \end{minipage}
\end{figure}

Drift velocity of electron and holes has been simulated for pPCGe detector as shown in the Fig.~\ref{drfit_electron_exp_sim} and Fig.~\ref{drfit_hole_exp_sim} respectively.
Maximum order of the velocity near the point contact is observed as $10^{7}$ cm$^{2}$/sec.
At the higher electric field (as moving towards point contact), the drift velocity of electrons as well hole has been found to be saturated as shown in the Fig.~\ref{drfit_electron_exp_sim} and Fig.~\ref{drfit_hole_exp_sim}  respectively and compared with the experimentally measured values \cite{ge_elect}\cite {ge_hole}. 
However the linear variation in the drift velocity has been also observed in the low electric field region.

\subsection{Pulse shape (charge signal and current signal)}
\begin{figure}[!tbp]
  \centering
  \begin{minipage}[b]{0.4\textwidth}
    \includegraphics[width=\textwidth]{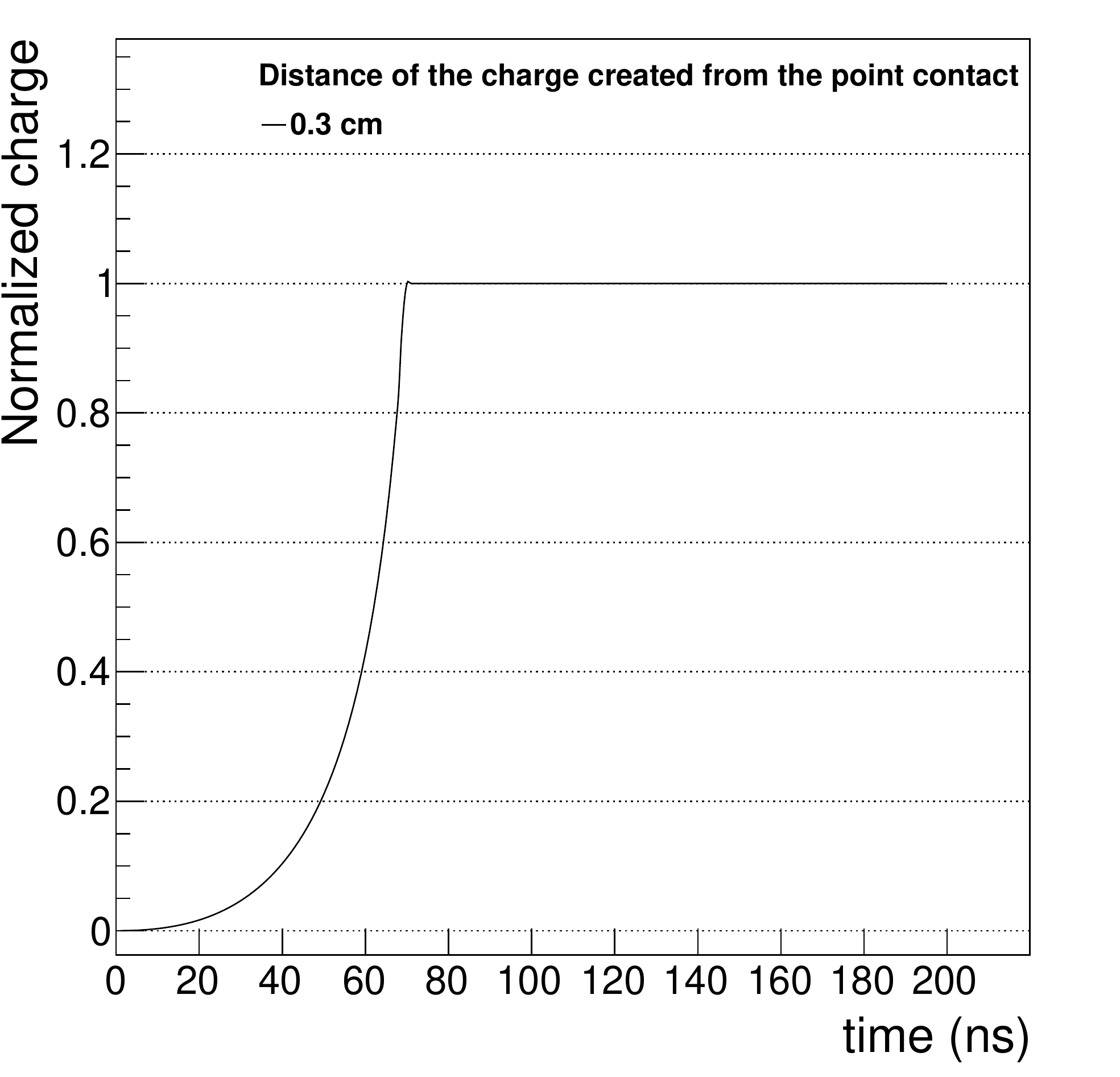}
    \caption{Normalized charge pulse of pPCGe detector}
    \label{induce_charge_point_three.pdf}
  \end{minipage}
  \hfill
  \begin{minipage}[b]{0.4\textwidth}
    \includegraphics[width=\textwidth]{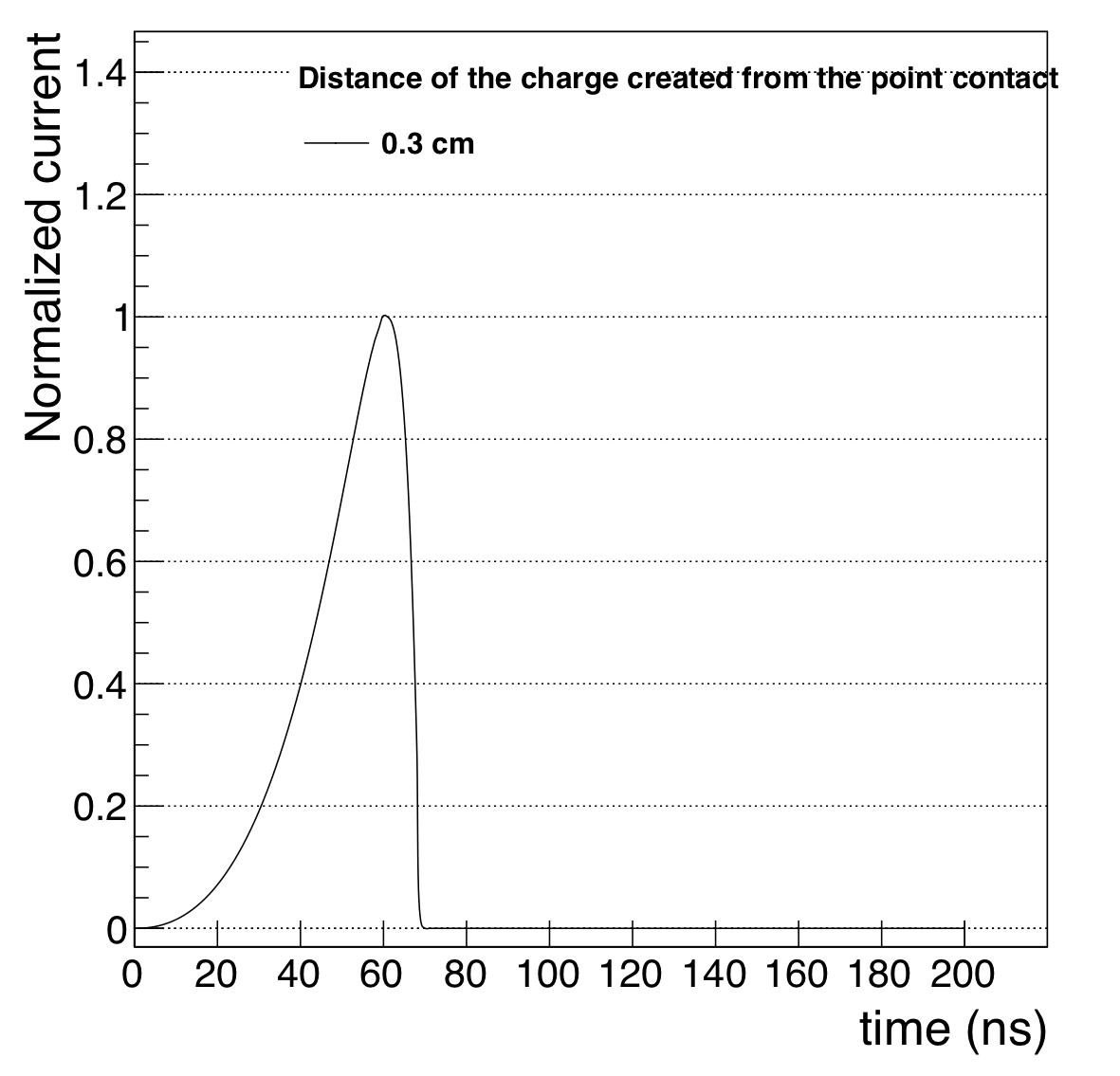}
    \caption{Normalized current pulse of pPCGe detector}
    \label{induce_current_point_three_copy}
  \end{minipage}
\end{figure}

\begin{figure}[!tbp]
  \centering
  \begin{minipage}[b]{0.45\textwidth}
    \includegraphics[width=\textwidth]{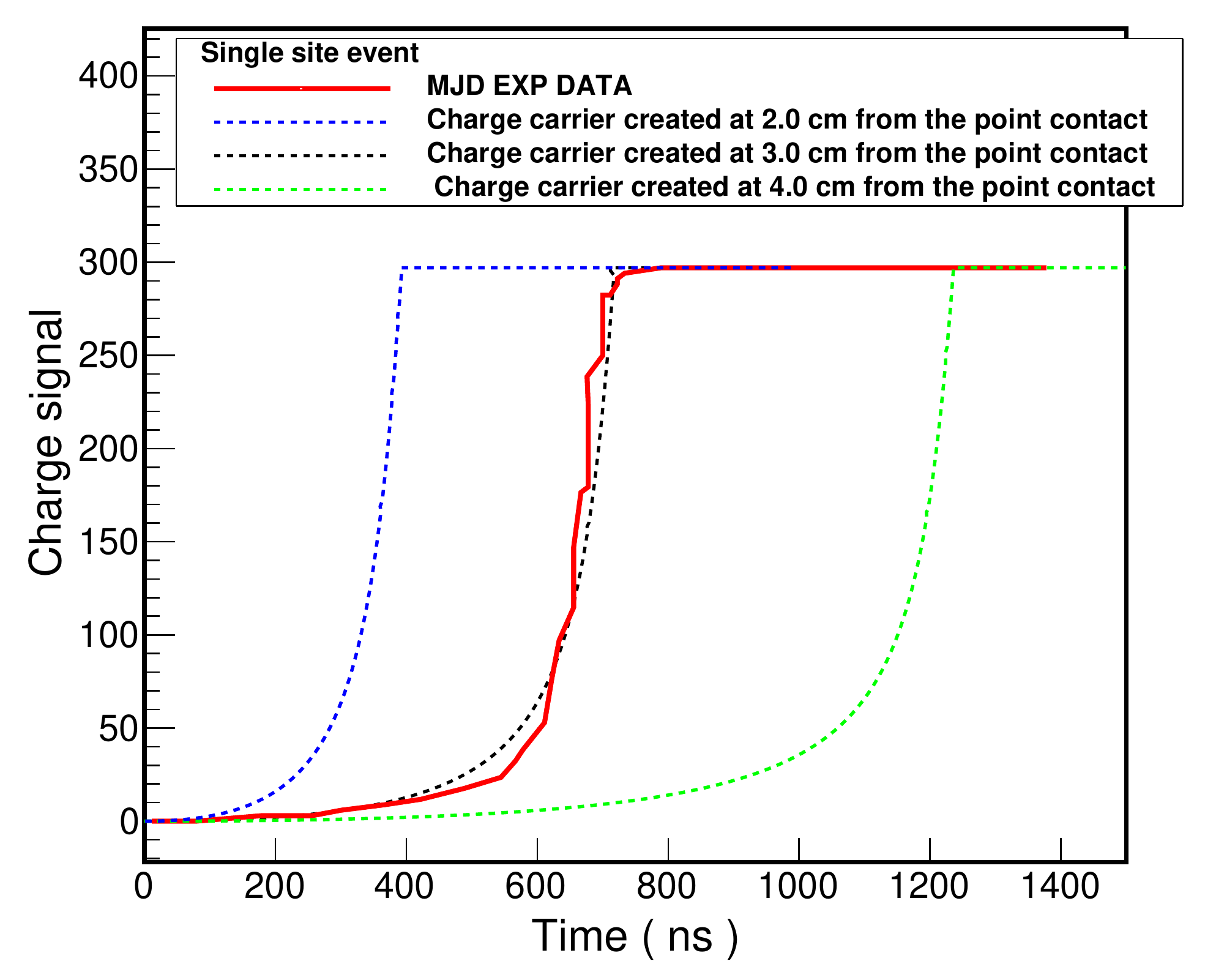}
    \caption{Simulated charge pulses for various initial positions of charge carrier within the pPCGe detector }
    \label{charge_signal.pdf}
  \end{minipage}
  \hfill
  \begin{minipage}[b]{0.45\textwidth}
    \includegraphics[width=\textwidth]{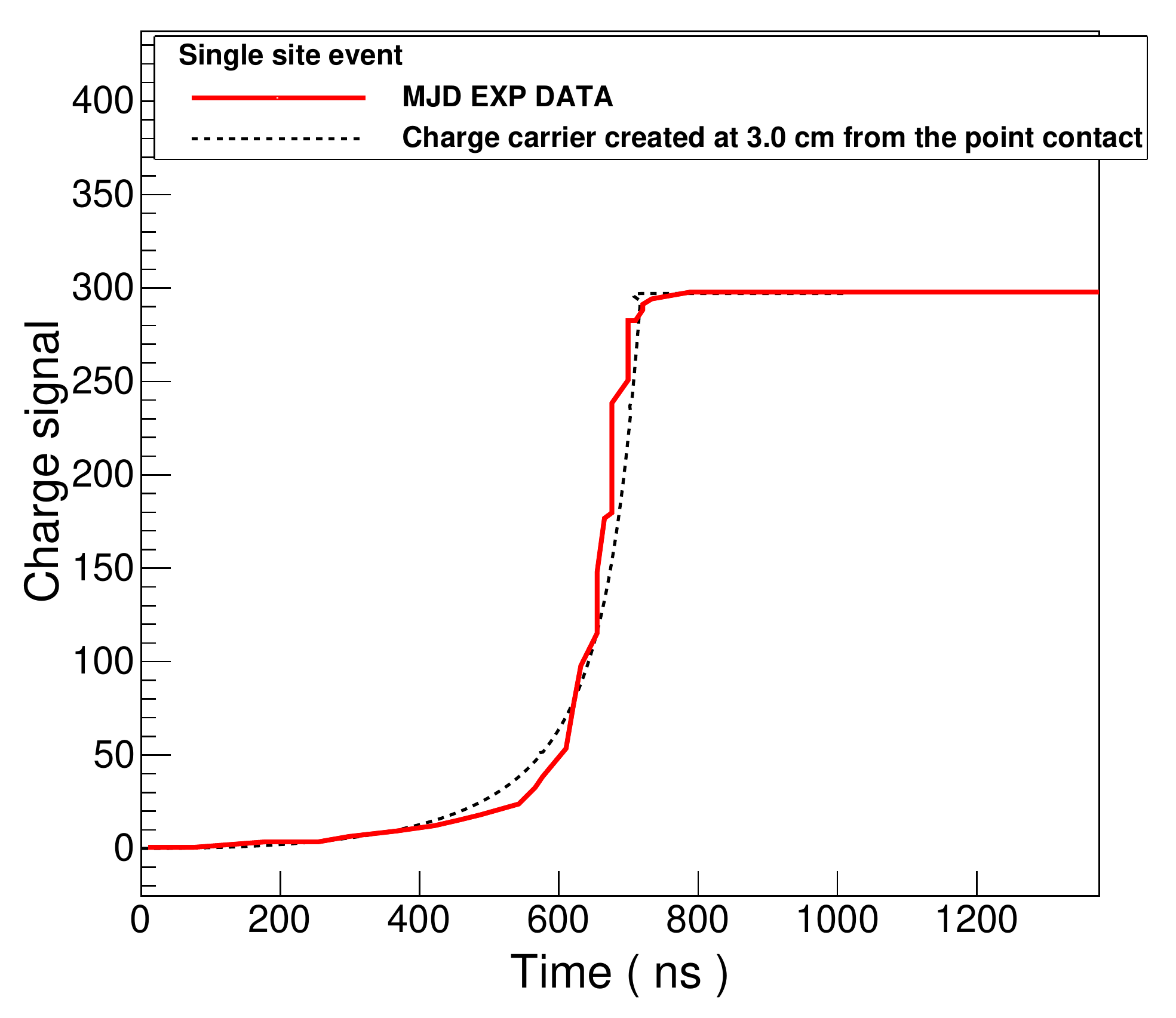}
    \caption{Simulated charge signal accordance with experimental charge pulse}
    \label{induce_current_point_three_copy.png}
  \end{minipage}
\end{figure}

Charge signal and current signal based on arbitrary detector parameters( Table$\sim$\ref{arb_det_par}) has been shown in the Fig.~\ref{induce_charge_point_three.pdf} and Fig.~\ref{induce_current_point_three_copy} respectively.
 The radiation which entered inside the detector volume gives rise to the energy deposition. As a consequence of energy deposition within the detector volume, charge carrier (electrons and holes) will produce due to the ionization.
 These charge carrier started drifting in the presence of applied electric field of the detector and induces the charge on the read out / electrode (point contact) of the detector . At every time instant of the charge carrier trajectory, charges will induce on the electrode and induction of charges continues till the charge carriers reaches the electrode. Therefore the charge signal will increases slowly as they moves towards the electrode. Once all charge carrier reaches the electrode, the charge signal will gets maximum height and achieve the charge saturation as shown in the Fig.~\ref{induce_charge_point_three.pdf} .
 Saturation of charge signal occurs because all the charges have been collected and detected on the electrode of the detector and as a consequence no further change has been observed in the charge signal irrespective of time.
 Current signal can be easily obtained from the time integration of charge signal as shown in the Fig.~\ref{induce_current_point_three_copy.png}.
 \begin{table}[h!]
\begin{center}
\setlength{\tabcolsep}{0.16em} 
\renewcommand{\arraystretch}{1.37}
\centering
\caption{Parameters used in the simulation based on the arbitrary detector attributes are tabulated  }
\vspace*{4pt}
\label{arb_det_par}
\footnotesize
\begin{tabular}{|c|c|} \hline
  
  Detector dimension         & 1 cm x 1 cm    \\ \hline  
   Grid dimension         & 200 x 200    \\ \hline  
  Point contact radius          & 0.5 mm                \\ \hline
Impurity concentration ( p-type Ge crystal )  & 5.0 x 10$^{10}$ cm$^{-3}$              \\    \hline
  Reverse bias voltage         &          3500 Volts         \\ \hline
 Gaurd ring dimension                     &  1 mm x 1 mm                   \\ \hline
 p+ ( boron diffused ) layer thickness                     &  300 nm                  \\ \hline
  \end{tabular}
\vspace{0mm}
\end{center}
\end{table}

\par
The charge signal of the pPCGe detector for Majorona Demonstrator (MJD) has been simulated based on the detector parameters~\cite{MJD} as given in the Table~\ref{MJD_par} . 
Simulated and the experimentally measured charge pulses have been plotted using the dotted and continuous lines respectively as shown in the  Fig.~\ref{charge_signal.pdf} and Fig.~\ref{induce_current_point_three_copy.png} .
Range of the y-axis (charge signal) has been chosen as [0,300] based on the MJD measured pulse~\cite{MJD} so that simulated and experimentally measured pulses can be compared.
Different initial position of charge carrier from the point contact has been fed into the simulation and the simulated charge signal has been shown in the Fig.~\ref{charge_signal.pdf}. It has been observed from the Fig.~\ref{charge_signal.pdf} that the charge signal, corresponds to the initial position (2.0 cm from the point contact) have been nearly superimposed with the experimentally Majorana Demonstrator pulse\cite{MJD}. Whereas the charge pulses simulated for the initial positions 3.0 cm and 4.0 cm are different from that of initial position 2.0 cm. As the initial position of charge carrier from the point contact decreases, the rise time of the signal decreases accordingly and as a consequence, the saturation of signal has been observed ahead of the time as shown in the Fig.~\ref{charge_signal.pdf} .

\begin{table}[h!]
\begin{center}
\setlength{\tabcolsep}{0.16em} 
\renewcommand{\arraystretch}{1.37}
\centering
\caption{pPCGe detector parameters used in the Majorona Demonstrator (MJD) are tabulated from Ref.~\cite{MJD}  }
\vspace*{4pt}
\label{MJD_par}
\footnotesize
\begin{tabular}{|c|c|} \hline
  
  Detector dimension         & 5 cm x 5 cm    \\ \hline  
   Grid dimension         & 200 x 200    \\ \hline  
  Point contact radius          & 2.5 mm                \\ \hline
Impurity concentration ( p-type Ge crystal )  & 3.2 x 10$^{9}$ cm$^{-3}$              \\    \hline
  Reverse bias voltage         &          3000 Volts         \\ \hline
 Gaurd ring dimension                     &  1 mm x 1 mm                   \\ \hline
 p+ ( boron diffused ) layer thickness                     &  300 nm                  \\ \hline
  \end{tabular}
\vspace{0mm}
\end{center}
\end{table}

\section{Summary and conclusion}
Electric field variation within the pPCGe detector has been simulated in the present simulation using the Finite Difference Method. As a consequence, the drift velocity profile of charge carrier inside the Ge crystal has also been figured out with the help of applied electric field and compared with the measured values.  It has been concluded from the simulation as well as the velocity saturation model that higher value of electric field is responsible for velocity saturation and the lower field region gives the linear variation of the drift velocity within the germanium crystal respectively. Weighting potential is one of the key input in the pulse shape simulation which has been simulated using the Shockley-Ramo Modeling.
Charge signal of pCGe detector has been simulated using the present simulation based on the Majorona Demonstrator (MJD) experiment . The simulation  reflects the fair agreement with the experimental results.
Various initial positions of charge carrier has been simulated and compared with the experimental pulse shape of Majorona Demonstrator (MJD) and it has been concluded that the charge signal, corresponds to the definite initial position of the charge carrier ( 3.0 cm from the point contact) matches with the experimental signal pulse. Therefore if the detector parameters ( such as impurity concentration, high voltage, detector size, point dimension) and the experimentally measured charge pulse are known then the initial distance of charge carrier, where it has been created, can be predicted with present pulse shape simulation.
 
\label{sec:summary}

\acknowledgments
 The authors  are thankful to the Ministry of 
Human Resource Development (MHRD), New Delhi, India for the financial support through 
Scheme for Promotion of Academic and Research Collaboration (SPARC) project 
No. SPARC/2018$-$2019/P242/SL. Authors are also thankful to the PIRE-GEMADARC COLLABORATION which is supported by NSF USA.

\end{document}